\documentclass[
 aip,
 jcp,
 amsmath,amssymb,
 reprint,
]{revtex4-1}

\usepackage{graphicx}
\usepackage{dcolumn}
\usepackage{bm}
\usepackage[utf8]{inputenc}
\usepackage[T1]{fontenc}
\usepackage{mathptmx}
\usepackage{etoolbox}

\makeatletter
\def\@email#1#2{%
 \endgroup
 \patchcmd{\titleblock@produce}
  {\frontmatter@RRAPformat}
  {\frontmatter@RRAPformat{\produce@RRAP{*#1\href{mailto:#2}{#2}}}\frontmatter@RRAPformat}
  {}{}
}%
\makeatother

\begin{document}

\title{Bottom-Up Prediction of Amorphous Poly(1,3-dioxolane) using \emph{Ab Initio} Reactive Machine-Learning Force Fields}

\author{Yizhi Song}
\author{Bharatha K. Gunawardana}
\author{Kriti Alam}
\author{Teresa Shah}
\author{Deepa Ranabhat}
\author{Hsin-Yu Ko}
\email{hsin-yu.ko@unt.edu}
\affiliation{Department of Chemistry, University of North Texas, Denton, Texas 76203, USA}

\date{\today}

\begin{abstract}
Poly(1,3-dioxolane) (pDXL) is a chemically recyclable polyether that can be synthesized by living cationic ring-opening polymerization with precise control of chain length into the ultra-high-molecular-weight (UHMW) regime, where it acquires enhanced mechanical properties. Like most polymers, however, it has amorphous condensed-phase structure, and diffraction measurements alone cannot resolve the atomistic microstructure that underlies them. In this work, we predict the ensemble of the pDXL microstructures by applying a recently developed AI-accelerated \emph{ab initio} bottom-up polymer structure prediction (AI$^{2}$-BPSP) framework that simulates the living polymerization of pDXL under experimental synthetic conditions with machine-learning force fields trained within van der Waals-corrected hybrid density functional theory. Partitioning the predicted X-ray and neutron diffraction signal between the growing chain and the surrounding monomer resolves its dominant feature into two counteracting modes: a monomer contribution near $q \approx 1.5$~\AA{}$^{-1}$ that decays as monomer is consumed, and a polymer contribution near $q \approx 1.6$~\AA{}$^{-1}$ that is absent in oligomers and grows with chain length. This work paves the way to direct accuracy validation of these statistically significant, first-principles, and chain-length resolved diffraction predictions against experiments.
\end{abstract}

\maketitle

\section{Introduction}

The global plastic waste crisis has created a pressing need for chemically recyclable polymers that can be depolymerized back to their constituent monomers and repolymerized when needed.\cite{coates_chemical_2020,rahimi_chemical_2017,hong_chemically_2017,jehanno_critical_2022} Among the chemically recyclable polymers, poly(1,3-dioxolane) (pDXL) is a particularly promising candidate.\cite{abel_chemically_2021} In the presence of strong acid catalysts, pDXL can be depolymerized, enabling a sustainable pathway to molecular-level renewal.\cite{plesch_aspects_1976,kubisa_degradation_2023} Beyond its chemical recyclability, pDXL is unique compared with other sustainable polymers since it can be precisely synthesized experimentally via living cationic ring-opening polymerization (CROP) of 1,3-dioxolane (DXL), yielding linear macromolecules of targeted chain length. Such precision control leads to low dispersity across a wide range of degrees of polymerization up to the ultra-high-molecular-weight (UHMW) regime, which provides favorable mechanical properties (e.g., toughness and ductility).\cite{abel_chemically_2021,hester_ultra-high-molecular-weight_2023,shen_easy_2024} The precision control also provides a means for molecular weight distribution (MWD)-based design of polymer properties.\cite{hu_designing_2025,gentekos_controlling_2019} These properties make pDXL an ideal starting point for downstream chemical modification toward targeted functionalities.

To enable the rational design of pDXL derivatives as a platform for sustainable and functional polymers with precision control, it is necessary to gain an atomistic understanding of the condensed-phase (micro-)structure of this macromolecule. However, pDXL (like most polymers) is amorphous, leading to statistical challenges in experimental characterization (e.g., via X-ray and neutron scattering) and typically limiting the resulting structural assignment to a one-dimensional radial distribution, which is averaged both angularly and over the conformational ensemble of the sample. Specific three-dimensional atomistic geometries therefore cannot be directly resolved from such experimental observations alone.\cite{arbe_insight_2020} This limited experimental access to atomistic structures in the amorphous phase directly impedes molecular-level understanding of the properties of pDXL and hinders rational chemical modification. To overcome this limitation, computer simulations are typically used to generate structural ensembles of amorphous polymers. Many pioneering simulation techniques for polymers have been proposed, e.g., polymer structure predictor (PSP),\cite{sahu_polymer_2022} polymer builder in CHARMM-GUI,\cite{choi_charmm-gui_2021} Markov state models,\cite{knoch_nonequilibrium_2017} and first-principles polymer structure prediction.\cite{huan_polymer_2020} However, these methods typically rely on prescribed polymer topology/folding, empirical potentials to describe the interatomic interactions,\cite{gartner_iii_modeling_2019,schmid_understanding_2023} and protocols that require system-dependent heuristics (e.g., melt-and-quench processes distinct from experimental synthesis, backmapping to bridge multiscale simulations,\cite{muller-plathe_coarse-graining_2002,dhamankar_chemically_2021} and the choice of lag time to ensure Markovianity).

To simplify polymer structure prediction, a recently developed AI-accelerated \emph{ab initio} bottom-up polymer structure prediction (AI$^{2}$-BPSP)\cite{ko_group_placeholder_2026} framework (see Sec.~\ref{sec:methods} for a brief summary) in our group directly simulates polymer synthesis with a Deep Potential (DP) machine-learning force field (MLFF)~\cite{zhang_deep_2018,zhang_end--end_2018} trained on first-principles interactions obtained from quantum electronic structure theory. By following an ensemble of independent reactive polymerization processes using AI$^2$-BPSP, we are able to: (1) eliminate the difficult prescription of macromolecular topology and initial folding with starting synthetic conditions, (2) avoid the need to introduce a proxy melt-and-quench process (which may or may not be applied experimentally), and (3) replace empirical potentials with first-principles quantum-mechanical interatomic interactions computed with the sophisticated van der Waals (vdW)-corrected hybrid density functional theory (DFT) for better chemical realism. Compared with pioneering applications of recently developed MLFFs\cite{deringer_machine_2019,behler_four_2021,unke_machine_2021} to polymer systems,\cite{hong_first-principles-based_2021,long_polymers_2024,simm_simpoly_2025,hooven_how_2026} AI$^{2}$-BPSP introduces reactive bond breaking and formation events, thereby allowing on-the-fly modification of polymer topology. In this work, we apply the AI$^{2}$-BPSP to predict the amorphous pDXL structure as a function of its chain length and determine the statistical mechanical (molecular dynamics) sampling needed to converge the predicted principal structural signals (X-ray and neutron scattering). Specifically, we report ensembles for pDXL microstructures from the dimer to the 100-mer, establish the ensemble size required for converged predictions, and present the X-ray and neutron diffraction signals computed from them with quantified uncertainty. Partitioning these signals into contributions from the growing chain and the surrounding monomers reveals that the dominant diffraction feature arises from two counteracting contributions. Taken together, this work opens the door to direct experimental validation using experimentally precisely synthesized pDXL over a wide range of chain lengths.

\section{Methods}\label{sec:methods}

The reactive MLFF used in this work was obtained using the SeA-enabled DP active-learning protocol introduced in Ref.~\onlinecite{gunawardana_first-principles_2026}. In this protocol, we generated an initial training dataset using ab initio molecular dynamics (AIMD)\cite{car_unified_1985,marx_ab_2009} at the PBE0+D3 level of theory driven by a recently developed SeA high-throughput hybrid-DFT engine.\cite{ko_high-throughput_2023} These initial data were then used to iteratively train a DP MLFF\cite{zhang_deep_2018,zhang_end--end_2018} through active-learning cycles.\cite{zhang_active_2019,zhang_dp-gen_2020} During each cycle, an ensemble of four equivalent DP MLFFs was trained (same data but different random seeds) using the DeePMD-kit package;\cite{wang_deepmd-kit_2018,zeng_deepmd-kit_2023} one of them was used to perform molecular dynamics exploration of the target thermodynamic conditions, while all four were used to evaluate the model deviation for selecting candidate configurations to be relabeled at the hybrid-DFT level and returned to the training set. Configurations whose model deviation fell in the window 0.20 to 0.35 eV/\AA{} were taken as candidates for relabeling. To account for the CROP processes for pDXL synthesis, we also included reactive MD exploration following a recent extension.\cite{ko_group_placeholder_2026} All training configurations were drawn from short oligomers solvated by monomers rather than from macromolecular pDXL chains. The monomer-solvated oligomers were deliberately placed in small periodic cells so that the nearby periodic images of the chain encode chain--chain interactions in the training data. Consequently, the short-ranged, oligomer-trained DP MLFF is transferable to the macromolecular limit as previously demonstrated using polyethylene.\cite{gunawardana_first-principles_2026} The model deviation over the production trajectories confirms that this holds for pDXL out to the 100-mer (Section S2 of the Supporting Information). For pDXL, the active-learning protocol converged to a training set of 40,359 labeled configurations of 144 atoms each at the PBE0+D3 level of theory. On held-out testing data the resulting force field gives a root-mean-square error of 83 meV/\AA{} for the force components and 2.7 meV/atom for the energies (Section S1 of the Supporting Information).

During this training process, all first-principles simulations were performed using the PWscf module of the open-source Quantum ESPRESSO\cite{giannozzi_quantum_2009,giannozzi_advanced_2017,giannozzi_quantum_2020} package (modified from version 7.2, Ref.~\onlinecite{ko_high-throughput_2023}) with the PBE0 hybrid functional\cite{perdew_rationale_1996,adamo_toward_1999} and Grimme's D3 dispersion correction with zero damping.\cite{grimme_consistent_2010} The core electrons were treated implicitly together with the nuclei via the ONCV norm-conserving pseudopotentials,\cite{hamann_optimized_2013,schlipf_optimization_2015} while the valence electrons were treated explicitly in a plane-wave basis with a kinetic energy cutoff of 120 Ry. The amorphous systems were modeled with periodic boundary conditions to minimize finite-size effects, and the first Brillouin zone was sampled at the $\Gamma$-point only. For simplicity, we treated the counterion of the cationic propagating chain implicitly as a uniform background charge compensating the total system charge of +1e.

\begin{figure}[tb]
  \centering
  \includegraphics[width=\columnwidth]{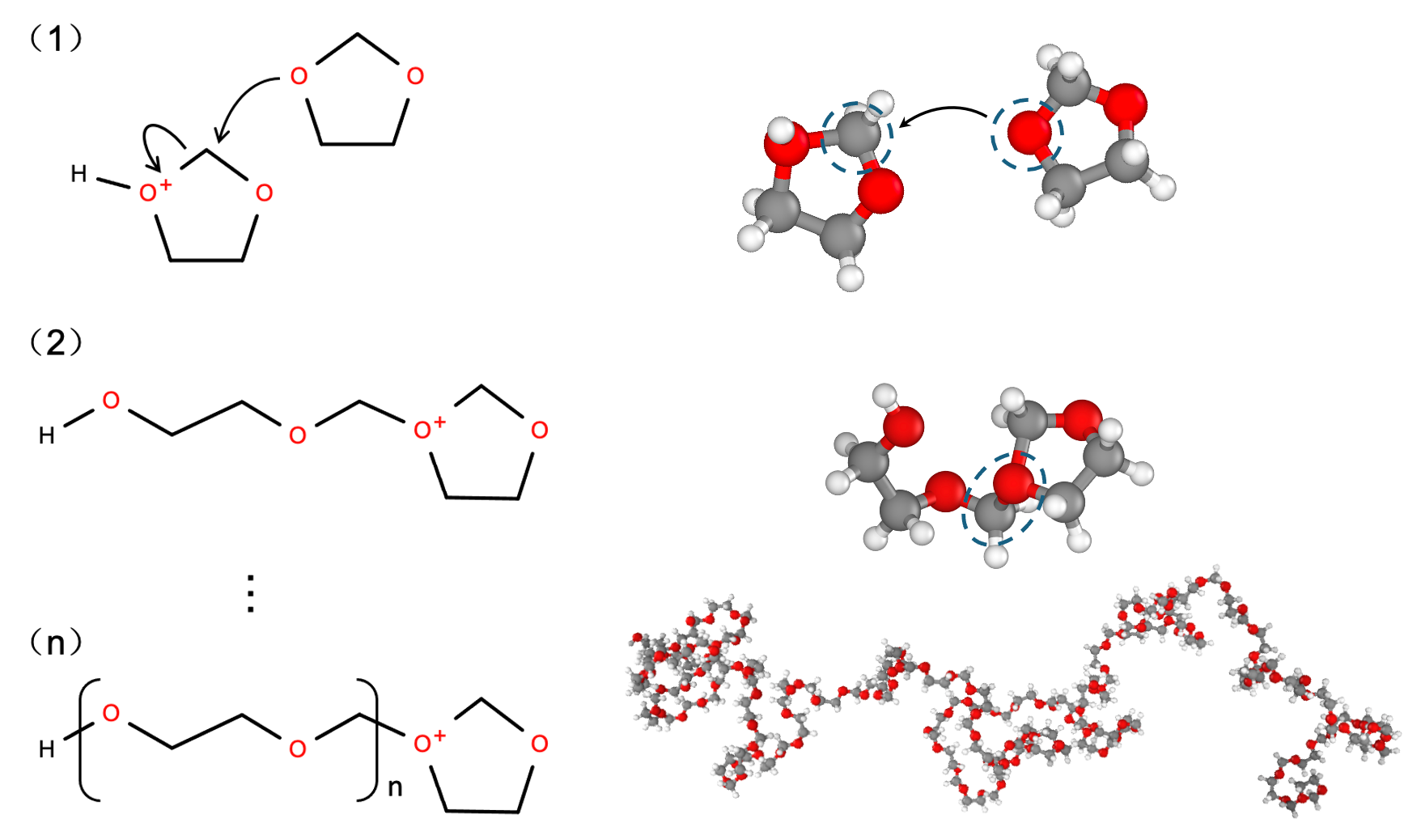}
  \caption{Cationic ring-opening polymerization of 1,3-dioxolane as simulated in this work. (1) An oxygen of an incoming DXL monomer attacks the carbon adjacent to the protonated chain end, breaking the C--O bond and opening the ring; the reacting atoms are circled in the ball-and-stick representation. (2) The resulting dimer carries the positive charge on the new terminal ring, so the same elementary step repeats. ($n$) Propagation to a degree of polymerization $n$. Left: Lewis structures, with curved arrows indicating electron flow. Right: corresponding configurations taken from the trajectories (C grey, O red, H white); the lowest panel shows a representative 100-mer. The unreacted DXL monomers that constitute the remainder of each simulation cell are omitted for clarity.}
  \label{fig:mechanism}
\end{figure}

Using the trained DP MLFF, we performed the AI$^{2}$-BPSP workflow\cite{ko_group_placeholder_2026} to simulate the polymer synthesis of pDXL. The key reaction modeled in this work is the CROP of DXL. In each step, an oxygen of a DXL monomer attacks a carbon at the chain end, breaking the C--O bond in the five-membered ring and incorporating the monomer into the growing chain (Figure~\ref{fig:mechanism}). Configurations along the ring-opening reaction coordinate of Figure~\ref{fig:mechanism} were added to the active-learning exploration, so that the force field is trained on the bond breaking and bond formation of the propagation step itself rather than on equilibrium conformers alone. Because pDXL synthesis proceeds experimentally through this single pathway via a living cationic mechanism,\cite{abel_chemically_2021} there is no need to introduce a kinetic model to account for competing/alternative reaction pathways.

\begin{figure*}[tb]
  \centering
  \includegraphics[width=\textwidth]{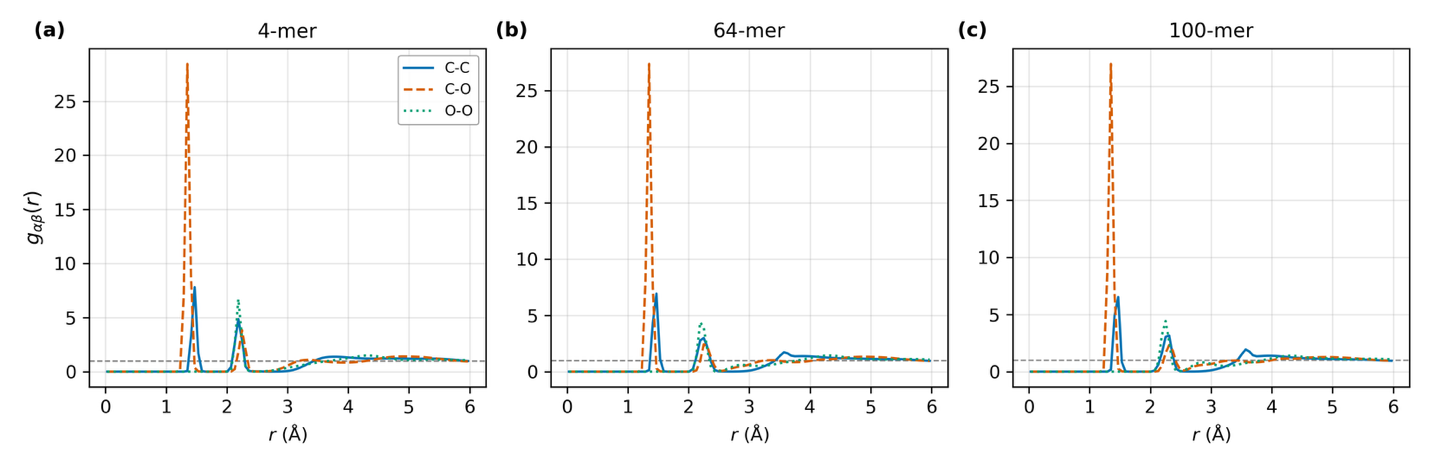}
  \caption{Heavy-atom (C--C, C--O and O--O) partial radial distribution functions $g_{\alpha\beta}(r)$ of the full system at $n = 4$, 64 and 100. The dashed line marks $g_{\alpha\beta} = 1$. Curves are averaged over 40 independent trajectories; shaded bands denote the standard error of the mean.}
  \label{fig:rdf}
\end{figure*}

Specific steps of our AI$^{2}$-BPSP workflow for pDXL are as follows. Starting from a monomer ($n = 1$), successive ring-opening events were performed with the reactive MLFF, with 20 ps of simulation per growth step, until 100-mer chains ($n = 100$) were obtained. At each growth step, the workflow identifies the active protonated chain end, associates a randomly chosen neighboring unreacted DXL monomer with it, applies a distance constraint between the attacking monomer oxygen and the chain-end carbon, and holds that association long enough for the local environment to reach thermal equilibrium; the ring-opening event itself is then carried out by the reactive MLFF, so no reaction pathway is imposed. Forty independent chain-growth trajectories were generated, yielding 4,000 distinct intermediate configurations spanning the entire polymerization process. Each periodic simulation cell contained 128 DXL molecules together with a single proton to initiate the cationic chain, giving 1409 atoms and a cell edge of approximately 23~\AA{}. The forty chain-growth trajectories started from the same initial configuration and were distinguished by the random seed applied at the first growth step, after which each followed its own sequence of ring-opening events. Independent production simulations were subsequently initialized from the distinct configurations generated at each degree of polymerization to ensure statistical independence of the structural ensemble. Each system was equilibrated and simulated for 100 ps in the NPT ensemble at 300 K and 1 bar using LAMMPS.\cite{plimpton_fast_1995,thompson_lammps_2022} Structural analysis was performed on the production trajectories at degrees of polymerization $n = 2$, 4, 8, 16, 32, 64 and 100.

From the resulting structural ensembles, we computed the partial structure factors in the Faber--Ziman convention,\cite{faber_theory_1965,liu_direct_2016}

\begin{equation}
S_{\alpha\beta}(q) = 1 + \frac{1}{x_{\alpha}x_{\beta}N}\left\langle \sum_{j=1}^{N_{\alpha}}\sum_{i \neq j}^{N_{\beta}} \exp\left[-i\mathbf{q}\cdot(\mathbf{r}_{j}-\mathbf{r}_{i})\right]\right\rangle,
\label{eq:fz}
\end{equation}
where $x_{\alpha}$ is the atomic fraction of species $\alpha$, $N_{\alpha}$ is the number of atoms of that species, and $r_{i}$ is the position of atom i. Angular brackets denote an average over the production trajectories. The scattering vector $q$ was sampled on the reciprocal lattice and spherically averaged over shells of equal $|q|$. For the polymer-only partials, the sums and the counts $N_{\alpha}$ run over the atoms of the growing chain only.

The X-ray interference function accessible to diffraction follows from the partials weighted by the $q$-dependent atomic form factors $f_{\alpha}(q)$,

\begin{equation}
F_{X}(q) = \frac{\sum_{\alpha}\sum_{\beta} x_{\alpha}x_{\beta}f_{\alpha}(q)f_{\beta}(q)\left[S_{\alpha\beta}(q)-1\right]}{\left[\sum_{\alpha}x_{\alpha}f_{\alpha}(q)\right]^{2}},
\label{eq:fx}
\end{equation}
where $f_{\alpha}(q)$ is defined as:

\begin{equation}
f_{\alpha}(q) = \sum_{i=1}^{4}a_{i}\exp\left[-b_{i}\left(\frac{q}{4\pi}\right)^{2}\right] + c.
\label{eq:cm}
\end{equation}
The coefficients $a_i$, $b_i$ and $c$ are the neutral-atom Cromer--Mann four-Gaussian parameters.\cite{cromer_x-ray_1968} The neutron interference function has the same structure, with the $q$-dependent form factors replaced by the $q$-independent coherent scattering lengths $b_{\alpha}$,\cite{sears_neutron_1992}

\begin{equation}
F_{N}(q) = \frac{\sum_{\alpha}\sum_{\beta} x_{\alpha}x_{\beta}b_{\alpha}b_{\beta}\left[S_{\alpha\beta}(q)-1\right]}{\left[\sum_{\alpha}x_{\alpha}b_{\alpha}\right]^{2}}.
\label{eq:fn}
\end{equation}
Both interference functions are defined so that $F(q) \rightarrow 0$ as $q \rightarrow \infty$. Because the form factors in Eq.~\ref{eq:fx} scale with atomic number, $F_X(q)$ is dominated by the C and O correlations, whereas the scattering lengths in Eq.~\ref{eq:fn} are comparable in magnitude across species, with that of H opposite in sign, so $F_N(q)$ is strongly sensitive to the hydrogen partials. Predicted spectra for the deuterated polymer were obtained by substituting $b_{\mathrm{D}}$ for $b_{\mathrm{H}}$ in Eq.~\ref{eq:fn} without modifying the underlying trajectories, so the hydrogenated and deuterated predictions derive from identical structural ensembles.

\begin{figure*}[tb]
  \centering
  \includegraphics[width=\textwidth]{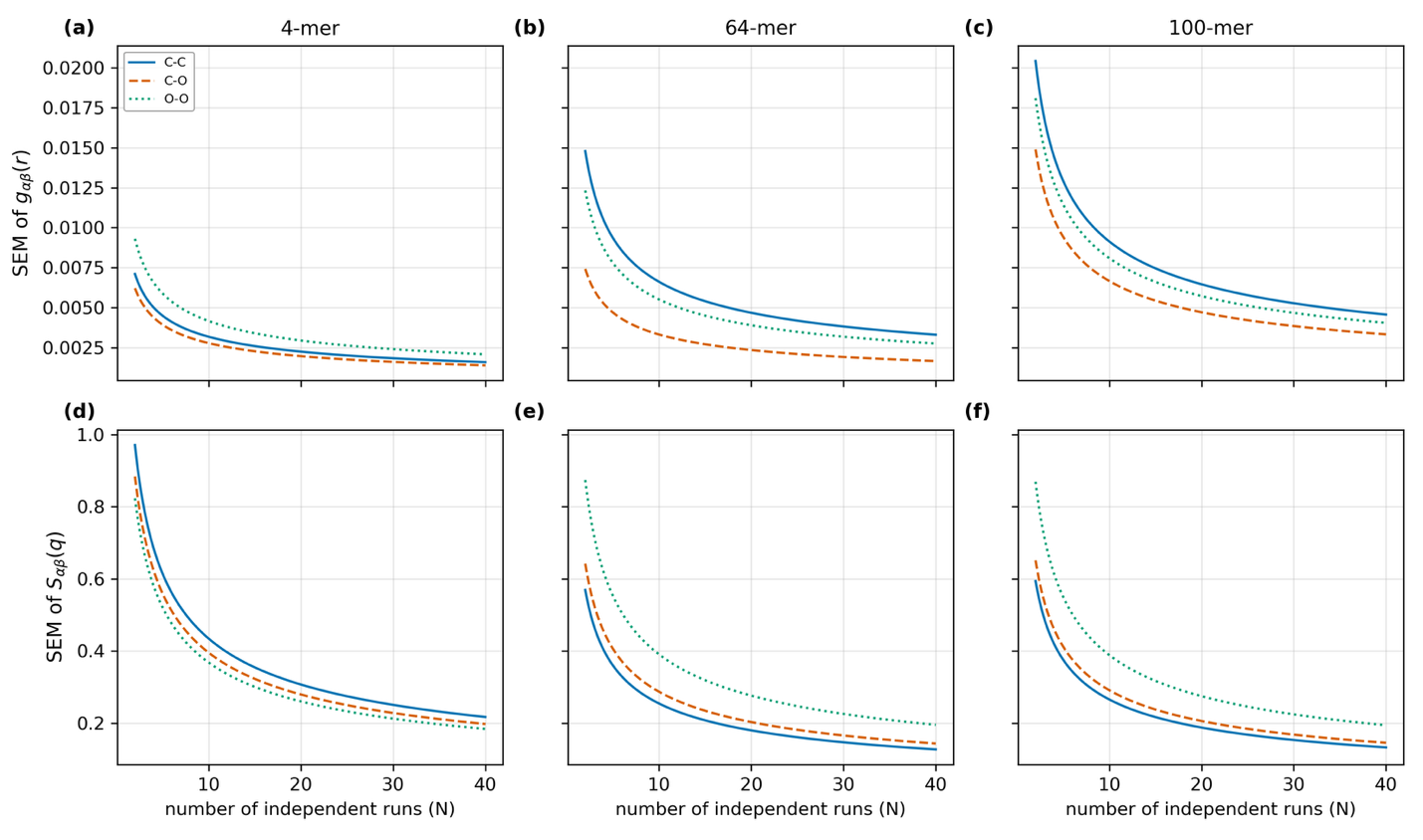}
  \caption{Statistical convergence of the predicted heavy-atom correlations with ensemble size. Standard error of the mean, $\sigma/\sqrt{N}$, for (a--c) the partial radial distribution functions \(g_{\alpha\beta}(r)\) and (d--f) the Faber--Ziman partial structure factors \(S_{\alpha\beta}(q)\), at degrees of polymerization $n = 4$, 64 and 100. $\sigma$ is the standard deviation across the 40 independent chain-growth trajectories, evaluated at each $r$ or $q$ and taken at its largest value within the analysis window defined in Section S3; $N$ is the number of independent trajectories. Curves are shown for the three heavy-atom pairs C--C, C--O and O--O, and the vertical scale is shared across the three panels of each row. The values quoted in the text correspond to $N$ = 40.}
  \label{fig:sem}
\end{figure*}

\section{Statistical Convergence of Polymer Structural Ensembles}

Bottom-up prediction through simulated polymer synthesis provides non-empirical sampling of macromolecular microstructures. During polymer synthesis, monomers can associate with the active growing site in various orders, necessitating independent sampling of multiple growth processes in simulation. In this section, we quantify the number of independent polymer-growth simulations needed to achieve statistical convergence of the underlying structural properties.

We began the assessment by examining 40 independent chain-growth simulations of pDXL with degrees of polymerization up to 100 and found that the structural properties are well converged, as shown by the radial distribution functions in Figure~\ref{fig:rdf} (see Section S8 and Figure S8 for the full assessment). Uncertainties are reported throughout as the standard error of the mean (SEM) over the independent chain-growth trajectories, $\sigma/\sqrt{N}$, where $\sigma$ is the sample standard deviation across trajectories at fixed $r$ or $q$; this is the quantity shown as shaded bands in the figures.\cite{flyvbjerg_error_1989,grossfield_quantifying_2009,grossfield_best_2018}

Every structural prediction reported below is an average over an ensemble of independently grown chains, so establishing the number of trajectories required for a converged average is a prerequisite for the rest of this work. Figure~\ref{fig:sem} reports the SEM of heavy-atom (C--C, C--O and O--O) partial radial distribution functions \(g_{\alpha\beta}(r)\) and the corresponding partial structure factors \(S_{\alpha\beta}(q)\ \)at $n = 4$, 64 and 100, as a function of the number of independent trajectories $N$. Each curve is $\sigma/\sqrt{N}$, where $\sigma$ is the standard deviation across the 40 independent trajectories at fixed $r$ or $q$, taken at its largest value within the analysis window defined in Section S3. The SEM decreases monotonically with ensemble size for every pair. At the full ensemble of $N = 40$, the worst-case SEM within the analysis window is 0.0046 for the RDF (C--C, 100-mer) and 0.22 for the partial structure factors (C--C, 4-mer), corresponding to 0.5\% and 3.4\% of the amplitude of the associated feature. SEM of \(g_{\alpha\beta}(r)\) and \(S_{\alpha\beta}(q)\) are plotted as shaded bands in Figure S8 and Figure~\ref{fig:partials}, respectively, where they are barely visible. Figure S3 shows that $\sigma$ itself is independent of ensemble size, which justifies scaling a single $\sigma$, taken from the full ensemble, as $\sigma/\sqrt{N}$.

\begin{figure*}[tb]
  \centering
  \includegraphics[width=\textwidth]{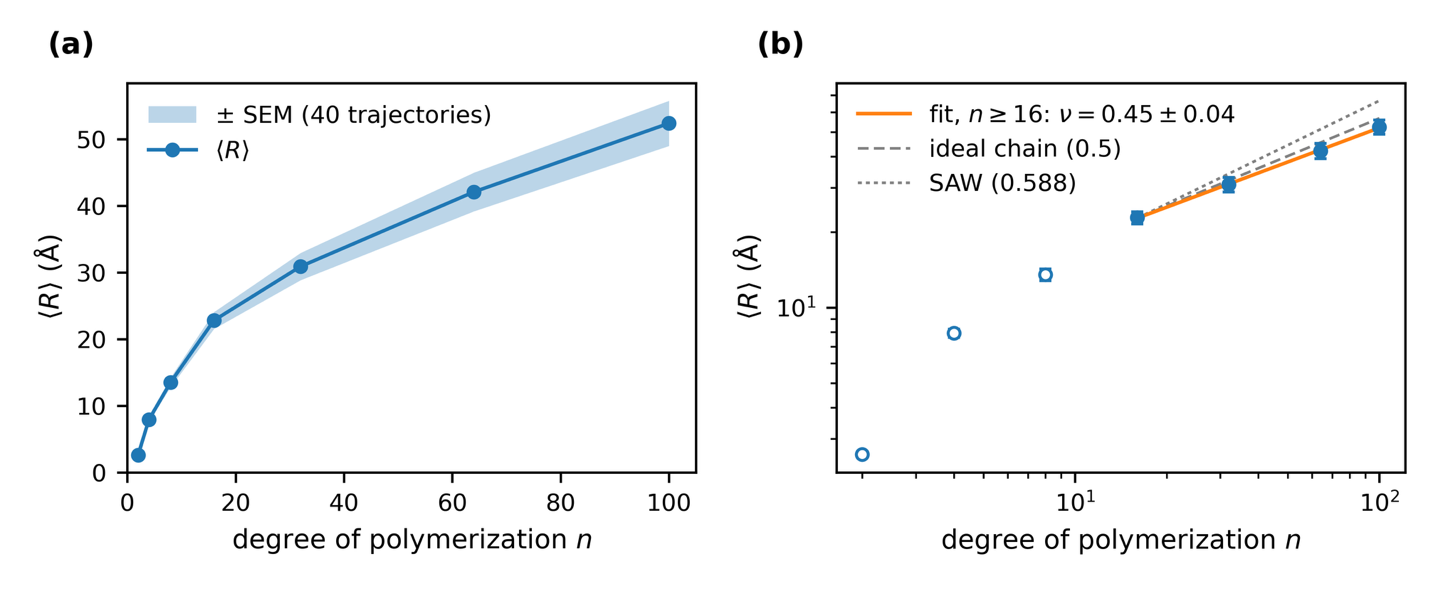}
  \caption{(a) Mean end-to-end distance $\langle R \rangle$ versus degree of polymerization $n$, averaged over 40 independent chain-growth trajectories; the shaded band is the standard error of the mean, $\sigma_{R}$/$\sqrt{40}$, where $\sigma_{R}$ is the standard deviation across trajectories of each trajectory's time-averaged end-to-end distance. (b) The same data on logarithmic axes, with a power-law fit giving $\langle R \rangle\sim n^{\nu}$ with $\nu$ = $0.45 \pm 0.04$; error bars are the standard error of the mean, and dashed and dotted lines show the ideal-chain ($\nu = 0.5$) and self-avoiding-walk ($\nu \approx 0.588$) slopes for reference. Filled symbols: included in the fit ($n \ge 16$); open symbols: shorter chains, shown but not fitted.}
  \label{fig:ree}
\end{figure*}

Figure~\ref{fig:ree}a shows the mean end-to-end distance $\langle R \rangle$ as a function of the degree of polymerization $n$, from 2-mer to 100-mer, with the shaded band denoting the standard error of the mean, $\sigma_{\langle R\rangle}$ = $\sigma_{R}$/$\sqrt{40}$, across the 40 independent chain-growth trajectories. At small $n$, $\langle R \rangle$ increases steeply, as short oligomers contain few backbone bonds and remain close to fully extended. The growth rate decreases progressively with $n$, indicating that longer chains adopt increasingly coiled conformations in which the chain folds back on itself. Beyond $n \approx 16$ the growth is consistent with a power law, $\langle R \rangle\sim n^{\nu}$, and a log--log fit over that range gives $\nu$ = $0.45 \pm 0.04$ (Figure~\ref{fig:ree}b), to be compared with $\nu = 1$ for a rigid rod, $\nu = 0.5$ for an ideal Gaussian chain, and $\nu \approx 0.588$ for a self-avoiding walk in good solvent.\cite{flory_principles_1953,li_critical_1995,rubinstein_polymer_2003} Because a single trajectory explores only a narrow range of end-to-end distances on the simulated time scale, the exponent characterizes the ensemble of independently grown conformations rather than a time-averaged equilibrium. The relative dispersion of the end-to-end distance, $\sigma_{R}$/$\langle R \rangle$, is independent of $n$ beyond $n \approx 16$, and agrees with the value expected for a Gaussian end-to-end vector\cite{rubinstein_polymer_2003} (Section S4 of the Supporting Information). Ideal-chain statistics therefore describe the predicted ensemble by two independent measures. The practical consequence is favorable: because the relative dispersion does not grow with molecular weight, an ensemble size calibrated on short chains remains adequate at $n = 100$, and the required number of trajectories does not increase with chain length. Taken together, these tests establish that the 40 independent trajectories used throughout this work are sufficient for the claims made below.

It is worth noting that the relaxation time of a 100-mer is far beyond the 100 ps production window, so a single trajectory cannot be considered as equilibrated, and the ensemble was not constructed based on this premise. These forty chains grow independently, with different starting points and different sequences of ring-opening events. Thus, they correspond to different long-range chain conformations. Each trajectory samples a local quasi-equilibrium near the conformation it has reached. All the reported values below are average over the ensemble.

\section{Non-Empirical Prediction of Experimentally Probeable Structural Signals}

\subsection{Directly Probeable Signals}

\begin{figure*}[tb]
  \centering
  \includegraphics[width=\textwidth]{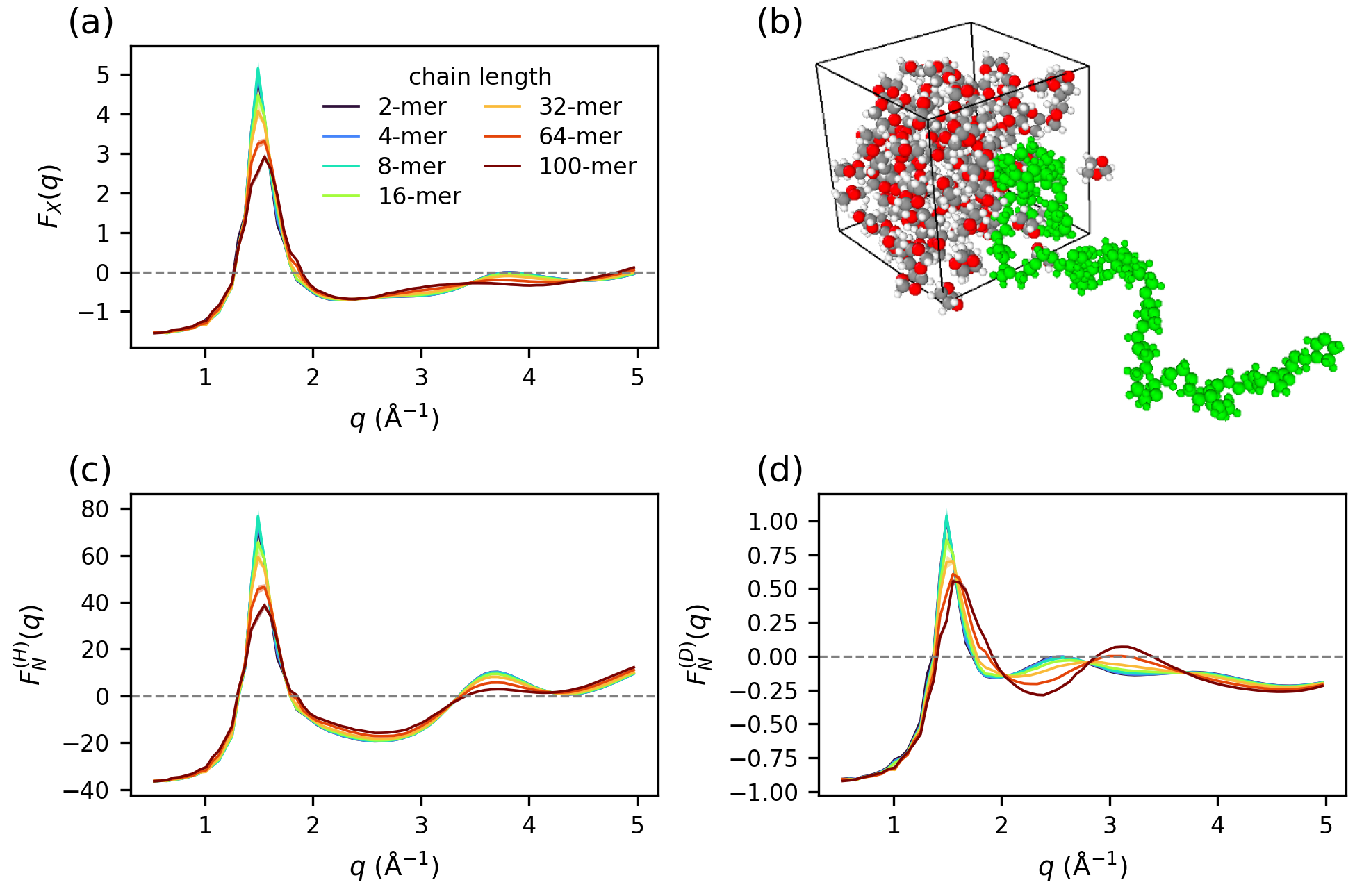}
  \caption{Predicted interference functions of the full simulation cell at degrees of polymerization $n = 2$, 4, 8, 16, 32, 64 and 100. (a) X-ray, $F_X(q)$. (b) Representative configuration of the simulation cell at $n = 32$, with the growing chain shown in green and the unreacted DXL monomers in red, white and grey. (c) Neutron for the fully hydrogenated polymer, $F_N^{(\mathrm{H})}(q)$. (d) Neutron for the fully deuterated polymer, $F_N^{(\mathrm{D})}(q)$. Curves are averaged over 40 independent chain-growth trajectories and shaded bands denote the standard error of the mean. Coherent scattering lengths are taken from Ref.~\onlinecite{sears_neutron_1992}.}
  \label{fig:ifull}
\end{figure*}

For most amorphous polymers, only a broad molecular-weight distribution is accessible, and the predicted and measured ensembles cannot be matched at a well-defined chain length; pDXL is an unusually favorable test case in this respect. Living CROP of DXL delivers chains of low dispersity across the full range of degrees of polymerization considered here,\cite{abel_chemically_2021,hester_ultra-high-molecular-weight_2023,shen_easy_2024} so a measured spectrum can be matched to a predicted spectrum at a known $n$. Here, we first predict the total X-ray and neutron interference functions of the complete system, which are the quantities accessible to a diffraction measurement. Figures~\ref{fig:ifull}a, \ref{fig:ifull}c and \ref{fig:ifull}d show $F_X(q)$, $F_N^{(\mathrm{H})}(q)$ and $F_N^{(\mathrm{D})}(q)$ for the full simulation cell across the chain-length series; representative configurations of the cell itself are shown in Figure~\ref{fig:ifull}b. Because these are the directly measurable quantities, they provide the foundation for assessing the accuracy of the predicted ensembles against experiment.

The three probes weight the same six partial structure factors differently. X-ray form factors scale with atomic number, so $F_X(q)$ is dominated by the carbon and oxygen correlations and is comparatively insensitive to hydrogen. In the hydrogenated neutron signal, the negative coherent scattering length of $^{1}$H opposes the positive values of carbon and oxygen, so the C--H and O--H cross terms enter with negative weight while the H--H term enters positively; this sign alternation amplifies the differences among the partials rather than averaging them, and produces a dynamic range an order of magnitude larger than that of $F_X(q)$. Deuteration renders all six weights positive and comparable in magnitude, so the weighted sum averages over the partials and the dynamic range contracts. Measuring both hydrogenated and deuterated samples therefore separates the heavy-atom and hydrogen contributions. The scattering lengths and the incoherent background of each sample are given in Section S5 of the Supporting Information.

Two features are common to all three probes. A primary peak at $q \approx 1.5$--1.6~\AA{}$^{-1}$ dominates every spectrum and carries the strongest chain-length dependence, decreasing in intensity from $n = 8$ to $n$ = 100. A secondary band spanning $q \approx 2.5$--4~\AA{}$^{-1}$ is weaker and evolves more slowly, appearing as a shallow negative lobe followed by a small positive feature near 3.7~\AA{}$^{-1}$ in $F_X(q)$, and as a pronounced negative excursion followed by a positive peak of comparable magnitude in $F_N^{(\mathrm{H})}(q)$. We expect these two principal bands to serve as the structural signals to be compared against experiment. As such, we investigated their structural origin in Secs.~\ref{sec:modes} and \ref{sec:chainlen}.

\begin{figure*}[tb]
  \centering
  \includegraphics[width=\textwidth]{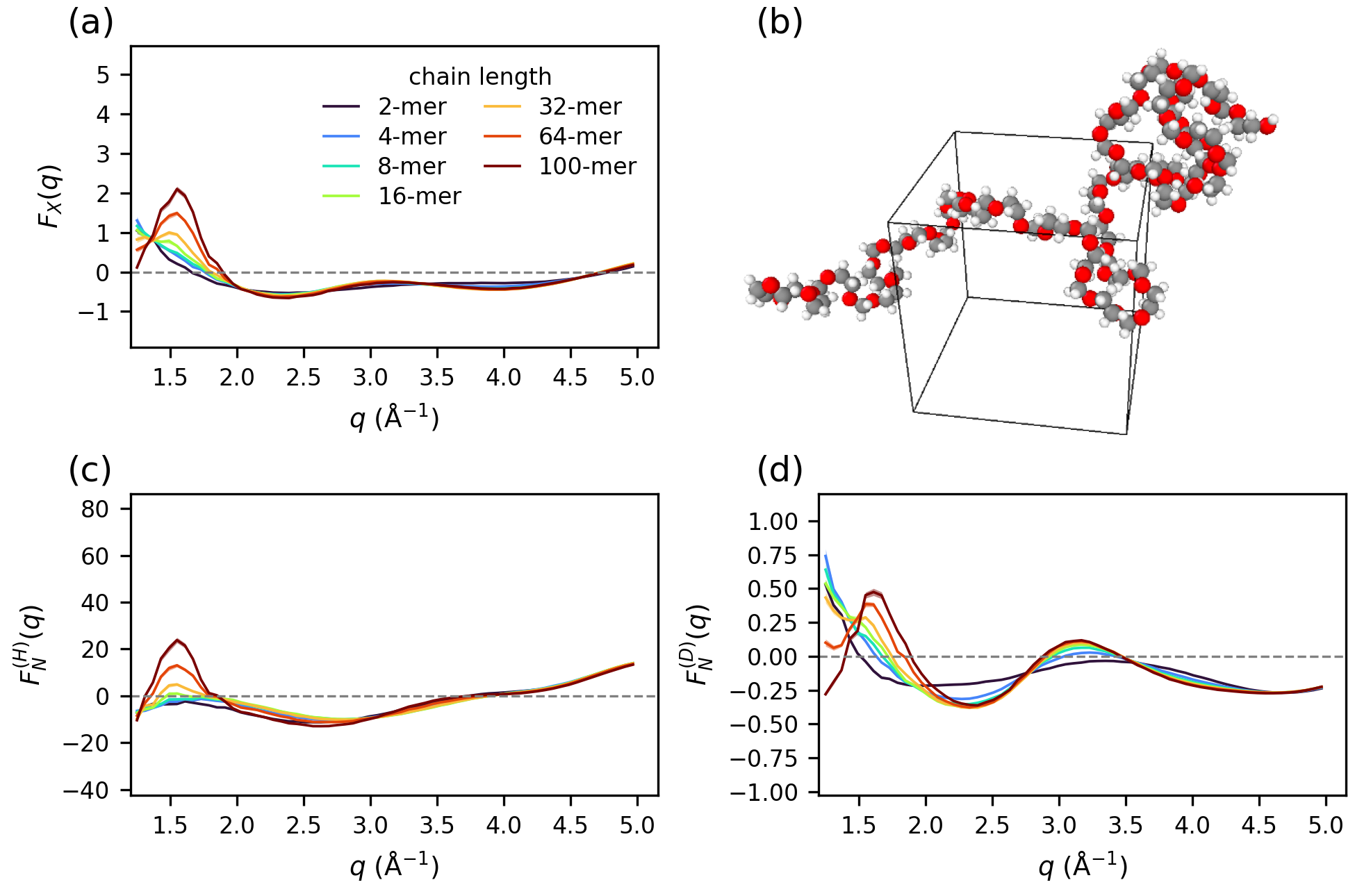}
  \caption{Predicted interference functions computed over the atoms of the growing chain alone: (a) X-ray, $F_X(q)$. (b) Representative configuration of the simulation cell at $n = 32$, shown unwrapped. (c) Neutron, hydrogenated. (d) Neutron, deuterated. Comparison with Figure~\ref{fig:ifull} separates the decaying monomer mode near $q \approx 1.5$~\AA{}$^{-1}$ from the growing polymer mode near $q \approx 1.6$~\AA{}$^{-1}$.}
  \label{fig:ichain}
\end{figure*}

\subsection{Understanding the Primary Chain-Length-Dependent Modes of the X-ray and Neutron Signals}\label{sec:modes}

The chain-length dependence of the primary peak is not a single trend. Because the solvent is unreacted DXL monomer, the composition of the cell changes systematically along the series, so the total signal is a composition-weighted sum over a polymer-monomer mixture whose ratio is set by the degree of polymerization. To separate the two contributions, we recomputed all three interference functions over the atoms of the growing chain alone (Figure~\ref{fig:ichain}; full $q$ range in Section S6, Figure S5). The partition resolves the primary peak into two modes of opposite chain-length dependence. A monomer mode centered near $q \approx 1.5$~\AA{}$^{-1}$ dominates the total signal while the chain is short and decays monotonically as the monomer is consumed. A polymer mode centered near $q \approx 1.6$~\AA{}$^{-1}$ is absent from the polymer-only spectra for $n \le 16$ and increases monotonically thereafter. The apparent attenuation of the primary peak in the full-system signal therefore does not indicate a loss of order within the polymer. It reflects the crossover between a decaying monomer mode and a growing polymer mode, and the modest shift of the peak to higher $q$ with increasing $n$ is the direct signature of that crossover. The secondary band at $q \approx 2.5$--4~\AA{}$^{-1}$ behaves differently. It is weak in the polymer-only spectra at every chain length, whereas in the full system it is present from $n = 2$ and diminishes as the monomer is consumed. The secondary band is therefore predominantly a feature of the monomer solvent rather than a property of the chain.

\subsection{Chain-Length Dependence of pDXL Structural Properties}\label{sec:chainlen}

\begin{figure*}[tb]
  \centering
  \includegraphics[width=\textwidth]{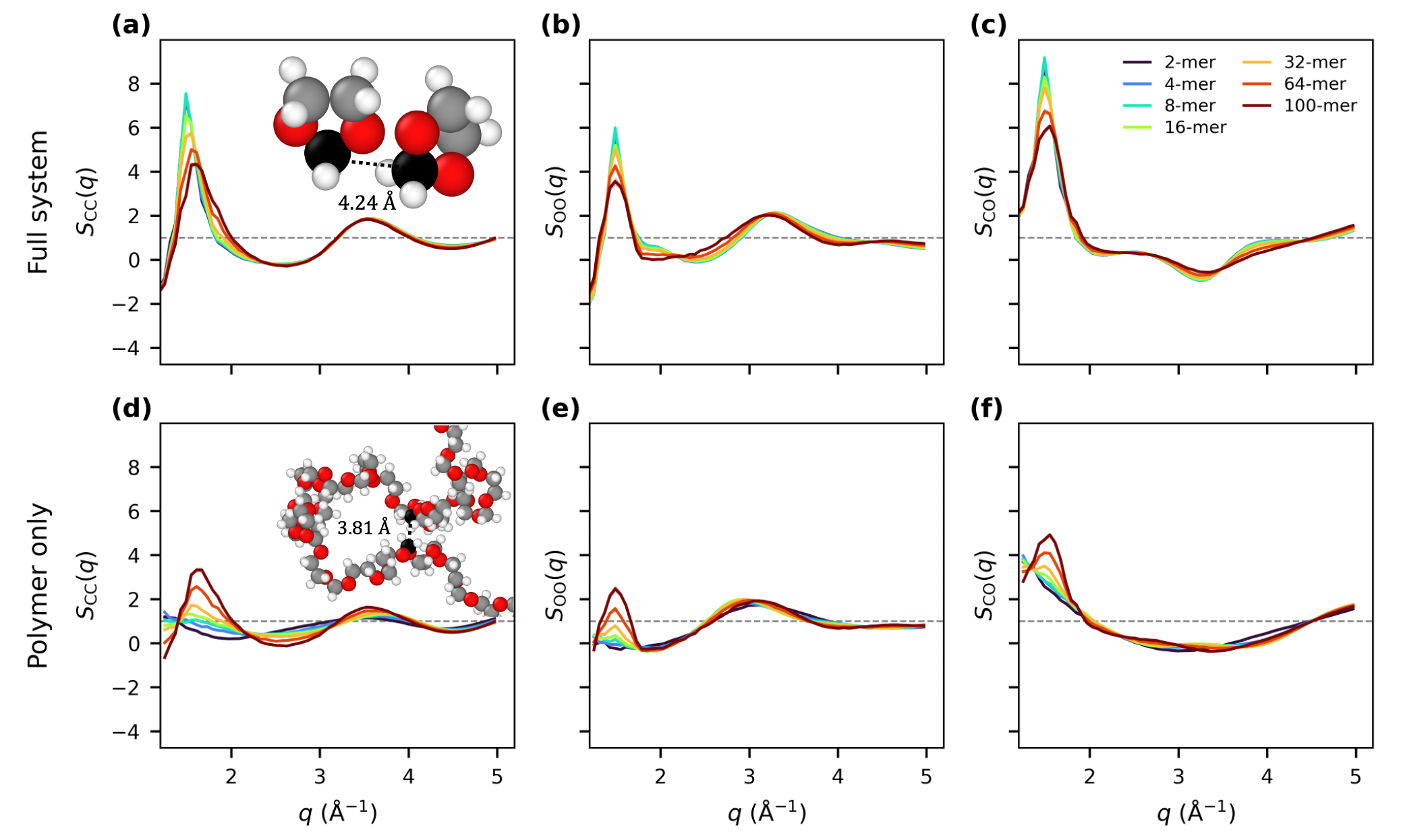}
  \caption{Faber--Ziman partial structure factors $S_{\alpha\beta}(q)$ of the pDXL system for the three heavy-atom pairs C--C, O--O and C--O, at degrees of polymerization $n = 2$, 4, 8, 16, 32, 64 and 100. Top row (a--c): full system with solvent. Bottom row (d--f): polymer atoms only. Curves are averaged over 40 independent chain-growth trajectories per chain length; shaded bands denote the standard error of the mean, and the dashed line marks $S_{\alpha\beta}$ = 1. All panels share the same y-axis range. Insets show the C$\cdots$C contacts responsible for the two modes, each taken from a single representative configuration: in (a), a monomer--monomer contact in the $n = 2$ system; in (d), an intrachain contact in the 100-mer between carbons separated by $\approx$ 15 repeat units along the backbone. The two carbons defining each contact are shown in black. Both modes are present in all three heavy-atom pairs; the contacts are illustrated for C--C, where they are largest.}
  \label{fig:partials}
\end{figure*}

Figure~\ref{fig:partials} shows the Faber--Ziman partial structure factors $S_{\alpha\beta}(q)$ for the three heavy-atom pairs (C--C, O--O, C--O), with the full system in the upper row and the growing chain alone in the lower row, on a vertical scale shared by all six panels (full $q$ range of the polymer-only partials in Figure S6). Both modes reside predominantly in the heavy-atom partials; the changes in hydrogen-containing partials are negligible (Section S7, Figure S7). In the full system, $S_{\mathrm{CC}}(q)$, $S_{\mathrm{CO}}(q)$ and $S_{\mathrm{OO}}(q)$ each show a pronounced principal peak near $q \approx 1.5$~\AA{}$^{-1}$, corresponding to a real-space correlation length of 4--4.5 \AA{} and reflecting the first intermolecular coordination shell of the monomer solvent (Figure~\ref{fig:partials}a inset; Figure S8). This peak attenuates beyond $n \approx 8$ while shifting to higher $q$. In the polymer-only partials, the same region is featureless at $n = 2$ and develops a peak near $q \approx 1.6$~\AA{}$^{-1}$ that grows monotonically with chain length and corresponds to non-bonded C$\cdots$C contacts at 3.4--4.0 \AA{} between chain segments separated by more than six bonds along the same backbone (Figure~\ref{fig:partials}d inset; Figure S9). Ring opening converts compact, orientationally uncorrelated DXL rings into extended chain segments, replacing the packing of the monomer solvent with a broader distribution of intrachain heavy-atom contacts; the monomer contribution is lost faster than the chain contribution is gained, so the total amplitude declines.

\section{Conclusions}

We have applied the AI$^{2}$-BPSP framework on amorphous pDXL to generate the chain-length-resolved ensembles of the underlying microstructures spanning degrees of polymerization from 2-mer to 100-mer. By gradually driving cationic ring-opening polymerization under NPT conditions at 300 K and 1 bar, we obtain amorphous structural ensembles that capture solvation and folding effects arising from explicit interactions with the condensed-phase environment, treated at the first-principles hybrid DFT+vdW level of theory. The computed partial structure factors provide direct predictions for X-ray and neutron scattering signals directly probeable from experiments. Partitioning them between the growing polymer chain and the monomer solvent shows that the primary peak is not a single feature but the sum of two counteracting modes: a decaying monomer contribution near $q \approx 1.5$~\AA{}$^{-1}$ and a growing polymer contribution near $q \approx 1.6$~\AA{}$^{-1}$. The apparent attenuation of the signal with chain length therefore reflects a crossover between them rather than a loss of order in the polymer. Intermediate-range order in amorphous pDXL emerges as the chain grows and is carried predominantly by the heavy-atom backbone alone; the hydrogen-containing correlations are essentially independent of chain length. The predicted contrast between hydrogenated and deuterated samples identifies a concrete measurement strategy: the two samples carry comparable fractional chain-length information at very different absolute contrasts, so a combined measurement would separate the heavy-atom and hydrogen contributions and provide a stringent test of the predicted ensembles. More broadly, these results show that bottom-up reactive simulation can supply diffraction-level structural predictions for an amorphous polymer without recourse to empirically constructed initial configurations.

\section*{Acknowledgments}

This work was supported by the University of North Texas (UNT). We acknowledge computational resources from the UNT CASCaM HPC cluster via National Science Foundation OAC-2117247 and the Texas Advanced Computing Center (TACC) at The University of Texas at Austin.

\section*{Author Declarations}

\section*{Conflict of Interest}

The authors have no conflicts to disclose.

\section*{Author Contributions}

Conceptualization: HYK

Methodology: YS, BKG, KA, TS, DR, HYK

Software: YS, HYK

Validation: YS, HYK

Formal analysis: YS

Investigation: YS

Data curation: YS

Writing --- original draft: YS

Writing --- review and editing: YS, HYK

Visualization: YS

Supervision: HYK

Funding acquisition: HYK

\section*{Data Availability}

The data that support the findings of this study are openly available in the
pdxl-dpmd-data repository on GitHub at
\url{https://github.com/YizhiSong/pdxl-dpmd-data} (version v1.0). The repository
contains the Deep Potential models used in all simulations, the initial
configurations of the 280 production simulations, the partial structure factors
and radial distribution functions computed for each simulation (full system and
polymer only), and the LAMMPS input file for the production runs.

\bibliography{references}

\end{document}